\documentclass[aip,jcp,
  asmath,amssymb,
  preprint,groupedaddress,showpacs]{revtex4}

\usepackage{graphicx}
\usepackage{epstopdf}
\usepackage{amssymb, amsmath}
\usepackage{multirow}
\usepackage{dcolumn}
\usepackage{bm}
\usepackage{subfigure}
\usepackage{color}
\usepackage{ulem}

\usepackage[final]{hyperref}
\hypersetup{
        colorlinks=true,       
        linkcolor=blue,          
        citecolor=blue,        
        filecolor=magenta,      
        urlcolor=blue
        }

\usepackage[utf8]{inputenc}
\usepackage[T1]{fontenc}
\usepackage{mathptmx}
\usepackage{etoolbox}

\usepackage{xr}

\newcommand\ket[1]     {|{{#1}}\rangle}
\newcommand\bra[1]     {\langle{{#1}}|}

\newcommand\Hop        {{\hat{H}}}
\newcommand\Kop        {{\hat{K}}}

\newcommand\Vop        {{\hat{V}}}

\newcommand\Order[1]   {\mathcal{O}\left(#1\right)}

\newcommand\Lz              {{\hat{L}_z}}

\newcommand\Jz              {{\hat{J}_z}}

\newcommand\LSterm[2]  {$^{#1} \textrm{{#2}}$}

\newcommand\eql[2] 
{
\begin{equation}\label{#1}
\begin{split}
#2
\end{split}
\end{equation}
}

\newcommand\eqsl[1]                            
{
\begin{align}
#1
\end{align}
}

\newcommand\CobaltIon {Co$^{2+}$ }
\newcommand\CoIIcomplex  {Co(C(SiMe$_2$ONaph)$_3$)$_2$ } 
\newcommand\CoIImodel {Co(C(SiMe$_2$OH)$_3$)$_2$ }

\newcommand\MnComplex {Mn$_{12}$ }

\newcommand\COMMENTED[1] {}

\makeatletter
\def\@email#1#2{%
 \endgroup
 \patchcmd{\titleblock@produce}
  {\frontmatter@RRAPformat}
  {\frontmatter@RRAPformat{\produce@RRAP{*#1\href{mailto:#2}{#2}}}\frontmatter@RRAPformat}
  {}{}
}%
\makeatother

\begin{document}
 
\title{Non-perturbative Many-Body Treatment of Molecular Magnets}

\author{Brandon Eskridge}
\email{bkeskridge@wm.edu}
\affiliation{Department of Physics, College of William and Mary, Williamsburg, Virginia 23187, USA}

\author{Henry Krakauer}
\affiliation{Department of Physics, College of William and Mary, Williamsburg, Virginia 23187, USA}

\author{Shiwei Zhang}
\affiliation{Center for Computational Quantum Physics, Flatiron Institute, New York, NY 10010, USA}
\affiliation{Department of Physics, College of William and Mary, Williamsburg, Virginia 23187, USA}
\email{szhang@flatironinstitute.org}

\begin{abstract}

Molecular magnets have received
significant attention because of  
their potential
applications in quantum information and quantum computing. A delicate
balance of electron correlation, spin-orbit coupling (SOC), ligand field
splitting, and other effects produces a persistent magnetic moment within
each molecular magnet unit. 
The discovery and design of molecular magnets with improved functionalities would be greatly aided by accurate computations.
However, the competition among the different 
effects poses a
challenge for theoretical treatments. Electron correlation plays a
central role, since $d$-, or $f$-element ions, which provide the magnetic states in
molecular magnets, often require explicit many-body treatments. 
SOC, which expands the dimensionality of the
Hilbert space, 
can also lead to non-perturbative effects in the presence of strong interaction.
Furthermore, molecular magnets are
large, with 
 tens of atoms in even the smallest systems.
We show how an \textit{ab initio} treatment
of molecular magnets can be achieved with auxiliary-field quantum Monte Carlo (AFQMC),
in which 
electron correlation, SOC, and material
specificity are included accurately and on an equal footing.
The approach is demonstrated by an application to compute the zero-field splitting 
of a locally-linear \CobaltIon complex.
\end{abstract}

\maketitle

\section{Introduction}

Molecular magnets
 were first experimentally realized 30 years ago with the magnetic characterization of a \MnComplex cluster which showed magnetic relaxation times on the order of two months at a temperature of 2K ~\cite{Mn12MolMag1993}.
Since then, 
broad classes of possible applications for 
molecular magnet systems have 
 been explored  
in areas such as high-density classical memory,
 quantum information~\cite{Vincent2012,PhysRevB.87.195412,TbPc2_electrically_driven,TbPc2_Grovers_2017,PerspectiveMolMag_Qinfo,Cr4_opticallyAddressable}, 
 and chemical catalysis~\cite{D2DT01440H}, among others.
For example, the TbPc$_2$ molecule has been used 
to implement Grover's algorithm within a single molecule~\cite{TbPc2_Grovers_2017}.
Since the original \MnComplex cluster,
many molecular magnets
(often referred to as single-molecule magnets in the literature)
 have been discovered and designed, based
 on 3d transition-metal~\cite{FeII_molMag_2010, family_of_FeII_molMags_2010, FeII_molMag_2011, CoII_molMag_2011, Linear_FeI_molMag_2013, FeII_molMag_2013, Co_molMag_2014, NiII_molMag_2015}, 
 lanthanide~\cite{TbPc2_Original_2003,doi:10.1021/jacs.6b02638,doi:10.1021/jacs.5b13584,DyMetallocene2018},
 and even actinide ions~\cite{ActinideMolMagRev2015, doi:10.1021/ja906012u, Neptunocene,Pu_MolMag_2018}.
While most molecular magnets 
display magnetic hysteresis only at temperatures below a few Kelvin, 
 a dysprosium metallocene cation 
 was recently discovered to display magnetic hysteresis at temperatures of up to 80 K~\cite{DyMetallocene2018}.
This represents an encouraging milestone for 
practical applications of molecular magnets in technology, since magnetic behavior occurs above the temperature of liquid nitrogen.
There is 
 significant interest in the design of new molecular magnets both for specific technological applications and fundamental science.

The defining characteristic of molecular magnets is the magnetic bistability that occurs due
 to their electronic structure.
The ground state of molecular magnets must 
be 2-fold degenerate, at least approximately, with non-zero total angular momentum,
 along with an energetic barrier that blocks spontaneous reversal of the magnetic moment.
The energetic barrier is provided by the 
 zero field splitting (ZFS) that 
 arises due to 
 a combination of
 spin-orbit coupling (SOC) and symmetry breaking from the ligand field.
 In the literature, ZFS is often used to refer to the parameters of the phenomenological pseudospin Hamiltonian
 used to model the effect;
 in the present work, however, we take ZFS to refer to the energy gaps in the low energy many-body spectrum
 of the \textit{ab initio} Hamiltonian.
Several relevant magnetic relaxation pathways exist for molecular magnets  
with phonon-mediated processes (Orbach, Raman, and direct electron-phonon scattering) and quantum tunneling of the magnetization 
 most often being the limiting factor in operating temperature.
Of course, a complete theoretical investigation of molecular magnets must consider these effects.
Much progress has been made in regards to designing efficient molecular magnets in terms of overall strategy for producing efficient molecular magnets based on the choice of magnetic ion, usually a $3d$-transition metal or lanthanide,
and ligand~\cite{WhatIsNotRequired4MolMags,C4CS00439F,Toward_HighT_SMMs,Review_molMag_2019}.
Accurate and reliable 
\textit{ab initio} treatments of molecular magnets 
would greatly facilitate designing molecular magnets and tuning their properties for specific use.

Molecular magnets pose challenges to explicit many-body treatments 
 due to the very large dimension of the Hilbert space necessary to describe them.
While exact solutions to the quantum many-electron problem are possible for small systems, the cost of exact methods scale exponentially in system size, which renders 
direct applications to typical molecular magnets 
not possible.
Approximate solutions based on density functional theory (DFT) are a natural choice; 
however, DFT may be inadequate due to the correlated $d$-, or $f$-element ions at the core of molecular magnets.
In general, 
Explicit many-body methods will most likely be needed for molecular magnets.
The quantum chemistry ``gold standard'' method, coupled cluster singles doubles with perturbative triples ( CCSD(T) ), scales as $N^7$ versus system size, $N$, making applications to typical molecular magnets challenging.
Additionally, while CCSD(T) reliably achieves chemical accuracy for main group chemistry, this is not always the case for 3d-transition metal chemistry~\cite{Hait2019,Shee2021}.

The typical challenges of performing many-body calculations are exacerbated by the inclusion of SOC which is fundamental to the computation of the ZFS.
The presence of spin-flip terms in the Hamiltonian expands the dimension of the Hilbert space that must be considered, greatly increasing 
the high computational cost of explicit many-body methods.
In much of the molecular magnet literature, the ZFS gaps and/or pseudospin Hamiltonian parameters are computed in two stages. 
First, static correlation is accounted for using a non-relativistic or scalar relativistic state-averaged CASSCF (SA-CASSCF)~\cite{SACASSCF_1980,SACASSCF_1981} calculation performed in an active space consisting of the magnetically active d-, or f-manifold.
Occasionally, a slightly larger active space is used which includes a few orbitals and electrons from the ligand as well.
In the SA-CASSCF calculations,  care is taken to average over the proper “no SOC” states since the specific states which are chosen for state averaging may influence the results.
Dynamic correlation is sometimes approximately accounted for using many-body perturbation theory, usually with 2nd-order N-electron valance perturbation theory (NEVPT2)~\cite{NEVPT2_2001,NEVPT2_2002} or 2nd-order complete active space perturbation theory (CASPT2)~\cite{CASPT2_1990,CASPT2_1991,msCASPT2_1998}. 
SOC is then treated 
 in a second stage via either quasi-degenerate perturbation theory (QDPT)~\cite{QDPT_2006}, or the restricted active space state interaction (RASSI) method~\cite{RASSI_2002}.
Such two stage approaches to the calculation of the ZFS have been remarkably successful for many 3d-transition metal, and lanthanide complexes.

In this work, we develop a general approach to treat molecular magnets using auxiliary-field quantum Monte Carlo (AFQMC) ~\cite{Zhang2003,AlSaidi2006b}. 
We recently incorporated explicit SOC in \textit{ab initio} AFQMC calculations which provides a 
computational framework where material specificity, both static and 
 dynamic correlations, and SOC are treated accurately and on an equal footing~\cite{Eskridge2022}. 
 AFQMC has demonstrated a high degree of accuracy in correlated electron systems in general, and systems containing 3d-transition metals specifically, as determined by several recent benchmarks~\cite{DirectCompMB2020, TM_bench_2019, TM_bench_2020, TM_metallocene_2022}, which, as discussed above, is an important factor for molecular magnets.
The calculation of the ZFS in molecular magnets
can be performed 
as a one-shot many-body calculation
 with no need to perform state-averaging, or to diagonalize the SOC operator in an explicit basis of many-body states.
AFQMC has a low order polynomial scaling versus system size (similar to  DFT but with a large prefactor), making applications to large systems feasible, even with the inclusion of SOC.
Still, the very large Hilbert space dimension of typical molecular magnets makes such applications very computationally demanding.
As an additional ingredient, we introduce
local embedding~\cite{Eskridge2019}, 
which produces an effective Hamiltonian in a basis of local orthonormal orbitals chosen based on local criteria,
to focus computational effort on 
magnetic ions while including much of the ligand as well.
The resulting Hamiltonian operates on a significantly reduced Hilbert space, which greatly increases the effective system size that can be 
treated. 
The accuracy of local embedding 
 AFQMC can be systematically improved towards full 
AFQMC treatment of the entire system 
by increasing the size of the local basis used. 

The remainder of the paper is organized as follows.
In Section~\ref{sec:theory}, we provide a brief summary of 
the general AFQMC framework, before describing
our approach for the non-perturbative treatment of molecular magnet
systems using AFQMC. 
In Section~\ref{sec:results}, we demonstrate the approach by applying it to compute the low-energy 
many-body spectrum, and ZFS, of the \CoIIcomplex molecule.
Comparisons are made with experimental results and other 
\textit{ab initio} many-body results from the literature~\cite{CoII_linear_complex}.
We conclude with some general remarks in Section~\ref{sec:Conclusion}.
In the supplemental material, we provide a Python script which reproduces 
the Hamiltonian used in Section~\ref{sec:results}.

\section{Theory}
\label{sec:theory}

In this Section, we describe the treatment of molecular magnets using AFQMC including explicit, non-perturbative SOC.
We focus specifically on the calculation of the low-energy many-body spectrum, and ZFS gaps.
In \textit{ab initio} AFQMC calculations, the many-body Hamiltonian is expressed in the second quantization formalism using a finite basis of orthonormal orbitals as may be obtained from self-consistent field (SCF) calculations.
A key component of performing efficient
 AFQMC calculations in molecular magnets 
is the production of an effective 
Hamiltonian
 which simplifies the rather complicated Hilbert space of the full system.
In Subsection~\ref{sec:molmag_treatment}, we describe a workflow for the non-perturbative treatment of molecular magnets, at the AFQMC level of theory, including a procedure to produce an interacting second quantized Hamiltonian.

\subsection{Auxiliary-Field Quantum Monte Carlo (AFQMC)}
\label{sec:AFQMC}

Here, we provide an overview of AFQMC~\cite{Zhang2003,AlSaidi2006b}.
A recent review outlines the general formalism in detail~\cite{Motta2018}, with a number of technical issues further discussed in Ref.~\cite{Shi2021}. 
AFQMC is an orbitally-based many-body method and is formulated in terms of a generic, interacting 
2nd quantized Hamiltonian,
\eql{eq:2ndQuantH}
{
\Hop = \Kop + \Vop =  \sum_{\mu \nu} K_{\mu \nu} \hat{c}^{\dagger}_\mu  \hat{c}_\nu +  \sum_{ \mu \nu \gamma \delta } V_{\mu \nu \gamma \delta} \hat{c}^{\dagger}_\mu  \hat{c}^{\dagger}_\nu \hat{c}_\delta \hat{c}_\gamma\,,
}
where $\Kop$ includes all one-body Hamiltonian terms, $\Vop$ includes all two-body interaction terms, $\hat{c}^{\dagger}_\mu$ and $\hat{c}_\mu$ are the fermionic creation and annihilation operators, respectively, which create/annihilate electrons in a chosen orthonormal basis of single-electron orbitals, and $K_{\mu \nu}$, $V_{\mu \nu \gamma \delta}$ are the matrix of elements of $\Kop$, $\Vop$ represented in the orbital basis.
Any standard form of the many-electron Hamiltonian can be represented by Eq.~\ref{eq:2ndQuantH}  including all-electron or pseudopotential Hamiltonians, and relativistic or non-relativistic treatments.

Observables are directly computed 
using a stochastic representation of many-body states in order to achieve high accuracy at a cost that scales as a low order polynomial.
The stochastic representation of a many-body wavefunction, $\ket{\Psi}$, is obtained via projection starting from an initial wavefunction, $\ket{\Psi_I}$, which has nonzero overlap with the target wavefunction.
The projection is performed in imaginary time as
\eql{eq:gs-proj}
{
    \lim_{\beta \to \infty}  e^{-\beta \Hop} \ket{\Psi_I} = 
    e^{-\tau \Hop}
    e^{-\tau \Hop}
    \cdots
    e^{-\tau \Hop}
    \ket{\Psi_I}
    \to
    \ket{\Psi}
    \,,
}
where the total projection time, $\beta$, has been divided into small imaginary time steps, $\tau$.

By Thoulesses' Theorem~\cite{THOULESS1960225}, the operation of the exponential of a one-body operator on a Slater determinant simply produces another Slater determinant; however, the presence of two-body Hamiltonian terms makes the projection nontrivial.
To handle general interactions, the projector, $e^{-\tau \Hop}$, is cast as a high-dimensional integral as follows.
First, the electron-electron interaction term is factored into a quadratic form of one-body operators, 
\eql{eq:Vquad}
{
\Vop = \sum_\gamma \hat{v}^2_\gamma 
\,,
}
where $\hat{v}_\gamma$ are one-body operators which are usually obtained from a modified Cholesky decomposition~\cite{Purwanto2011}.
The set of $\hat{v}_\gamma$ are truncated based on a small cutoff threshold, $\delta_{Chol.}$, which introduces 
 a systematically improvable approximation.
Second, a Trotter-Suzuki~\cite{Trotter1959,Suzuki1976} decomposition is performed,
\eql{eq:Trotter-Suzuki}
{
    e^{-\tau \Hop}
    \approx
    e^{-\tau \Kop / 2}
    e^{-\tau \sum_\gamma \hat{v}^2_\gamma }
    e^{-\tau \Kop / 2}
    + \Order{\tau^3}
    \,,
}
followed by a  Hubbard–Stratonovich transformation~\cite{Stratonovich1957,Hubbard1959} of $e^{-\tau \sum_\gamma \hat{v}^2_\gamma }$.
The projector is then given by,
\eql{eq:HSfull}
{
    e^{-\tau \Hop}
    \approx
      \int d\bm{\sigma} P(\bm{\sigma}) B(\bm{\sigma})    + \Order{\tau^3}
\,,
}
where $\bm{\sigma}$ is a vector containing auxiliary-fields, $P(\bm{\sigma})$ is a normal distribution function,  and $B(\bm{\sigma})$ is given by
\eql{eq:HSprojector}
{
B(\bm{\sigma}) =   e^{-\tau \Kop / 2} e^{\sqrt{\tau} \bm{\sigma} \cdot {\hat{\mathbf v}}}  e^{-\tau \Kop / 2}
\,, 
}
where $\hat{\mathbf v}$ is the vector of one-body operators defined implicitly by Eq.~\ref{eq:Vquad}.
Since $B(\bm{\sigma})$ is simply a product of exponentials of one-body operators, it can be easily applied to Slater determinants and the integral is evaluated by sampling auxiliary-fields from $P(\bm{\sigma})$ at each imaginary time step.
Many-body states are represented in an over-complete basis of non-orthogonal Slater determinant random walkers as
\eql{eq:mcPsi}
{
\ket{\Psi} \doteq \sum_k \ket{\Phi_k}
\,,
}
where $\ket{\Phi_k}$ is a single Slater determinant random walker, and $k$ runs over all walkers.
Thus, the projection has been cast as a random walk in nonorthogonal Slater determinant space.

The sign/phase problem, which is a generic problem that effects all fermionic quantum Monte Carlo approaches,
is controlled by the phaseless approximation in \textit{ab initio} AFQMC calculations~\cite{Zhang2003}.
The phase problem arises from the fact that physical observables are invariant under an arbitrary complex phase of the wavefunction from which they are computed.
As the random walk progresses, walkers accumulate a random phase relative to each other due to the, generally, complex-valued projector.
An importance sampling transformation based on the overlap of individual walkers with a trial wavefunction, $\ket{\Psi_T}$, is 
used to cast the random walks in a fixed gauge choice, 
which provides the theoretical basis to control this problem~\cite{Zhang2019}.
The trial wavefunction is an approximation to the target many-body state, and must have non-zero overlap with the desired exact many-body state.
With importance sampling, the Monte Carlo representation of the many-body wavefunction becomes,
\eql{eq:mcPsi_importance}
{
\ket{\Psi} \doteq \sum_k w_k \ket{\Phi_k}
\,,
}
where $w_k$ is a weight which is accumulated over the course of the random walk based on a chosen importance function, $\textrm{I}$, at projection step $n$ as $w^{(n)}_k = \textrm{I}(\bm{\sigma}, \Phi^{(n-1)}_k) w^{(n-1)}_{k} $.
Random walkers are still free to diffuse across the entire complex plane defined by $\bra{\Psi_T } \Phi_k \rangle$, allowing a finite density of walkers to accumulate at the origin, causing walker weights to diverge as the walk progresses.
The phase problem is then eliminated by projecting each individual walker onto an evolving line in the complex plane.
This is achieved by multiplying each walker by $\textrm{max}\{0, cos(\Delta \theta)\}$, where $\Delta \theta$ is the phase of $\bra{ \Psi_T} \Phi_k \rangle / \bra {\Psi_T} \Phi_{k-1} \rangle$.
The phaseless approach introduces a bias which can be 
controlled by the quality of the trial wavefunction. 

The most straight-forward application of AFQMC is the computation of the ground state energy.
Excited state AFQMC calculations are possible 
if a suitable trial wavefunction is used~\cite{Purwanto2009_C2}.
This typically requires a multideterminant trial wavefunction as may be obtained from CASSCF or other approaches.
Other more advanced projection methods are possible \cite{Ma_2013} but we will limit ourselves to the more conventional approach 
with multi-determinants. 

SOC can be treated explicitly in \textit{ab initio}   AFQMC calculations~\cite{Eskridge2022} 
since the SOC term, $\hat{K}^{soc} \equiv \hat{\mathbf{W}}^{soc} \cdot \hat{\mathbf{S}}$, is of a general one-body form as in generalized Hartree-Fock.
Several choices of $\hat{\mathbf{W}}^{soc}$ are possible ranging from all-electron relativistic Hamiltonians 
to formally non-relativistic model Hamiltonians based on relativistic pseudopotentials (PSPs) or effective core potentials (ECPs).
Explicitly, the 2nd-quantized Hamiltonian with SOC in a spin-orbital basis is given by 
\eql{eq:embedding_soc_H}
{
\hat{H}^{soc}  = \sum_{\mu \nu} \left( K_{\mu \nu} + K^{soc}_{\mu \nu} \right) \hat{c}^\dagger_\mu \hat{c}_\nu
  + \sum_{\mu \nu \gamma \delta} V_{\mu \nu \gamma \delta}  \hat{c}^\dagger_\mu \hat{c}^\dagger_\nu \hat{c}_\delta \hat{c}_\gamma
 \,,
}
where
$K^{soc}_{\mu \nu} = [ W^{z} S_{z} + W^{+} S_{+} +  W^{-} S_{-}]_{\mu \nu}$.
 The greek indices, $\mu$, $\nu$, $\gamma$, $\delta$ correspond to 
 spin-orbitals of the form $\chi_{\mu=(i,\sigma)} = \phi_i (\vec{r}) \ket{\sigma}$, 
where $\phi_i (\vec{r})$ are spatial orbitals, and $ \ket{\sigma}$ are eigenstates of the single-particle $\hat{s}_z$ operator.

The AFQMC procedure is
formally
 unchanged by the inclusion of explicit SOC; however, a few practical adaptations must be made.
With no SOC, the HS propagator, $B(\mathbf{\sigma})$, of Eq~\ref{eq:HSprojector} 
can be factorized 
\eql{eq:HSproj_factored}
{
 B(\mathbf{\sigma}) = B^\uparrow (\mathbf{\sigma}) \otimes  B^\downarrow(\mathbf{\sigma})\,,
}
where $B^\uparrow (\mathbf{\sigma})$ ($B^\downarrow (\mathbf{\sigma}) $) operates only on the up (down) spin sector, and 
Slater determinant random walkers are given as 
\eql{eq:SD_factored}
{
 \ket{\Phi_i} = \ket{\Phi^\uparrow_i} \otimes \ket{\Phi^\downarrow_i} \,.
}
With SOC, the HS projector explicitly mixes spins and has the form
\eql{eq:gBop}
{
B^G(\bm{\sigma}) =
  \begin{bmatrix}
   B^{\uparrow} (\bm{\sigma})&
   B^{+} (\bm{\sigma}) \\
   B^{-}  (\bm{\sigma})&
   B^{\downarrow} (\bm{\sigma})
   \end{bmatrix}
\,,
}
where the spin-flip  propagators $B^{+/-}  (\bm{\sigma}) = \mathrm{Exp}[-\tau \hat{W}^{+/-} ]$ are Hermitian conjugates of each other,
and $B^{\uparrow} (\bm{\sigma})$ ($B^{\downarrow} (\bm{\sigma})$) includes both the usual spin-independent terms, and the z-projection of the SOC term.
Generalized 
Slater determinant random walkers 
\eql{eq:gDet}
{
\Phi^G =
  \begin{bmatrix}
   \Phi^{\uparrow \uparrow} &
    \Phi^{\uparrow \downarrow} \\
   \Phi^{ \downarrow \uparrow}  &
   \Phi^{\downarrow \downarrow}
   \end{bmatrix}
\,,
}
are used 
during the AFQMC projection.
Thus, SOC is included exactly and on an equal footing with electron correlation.
The effective system size is doubled compared with treatments that neglect SOC.

\subsection{Treatment of molecular magnets}
\label{sec:molmag_treatment}

\begin{figure}
\begin{center}
\includegraphics[width=0.75\textwidth]{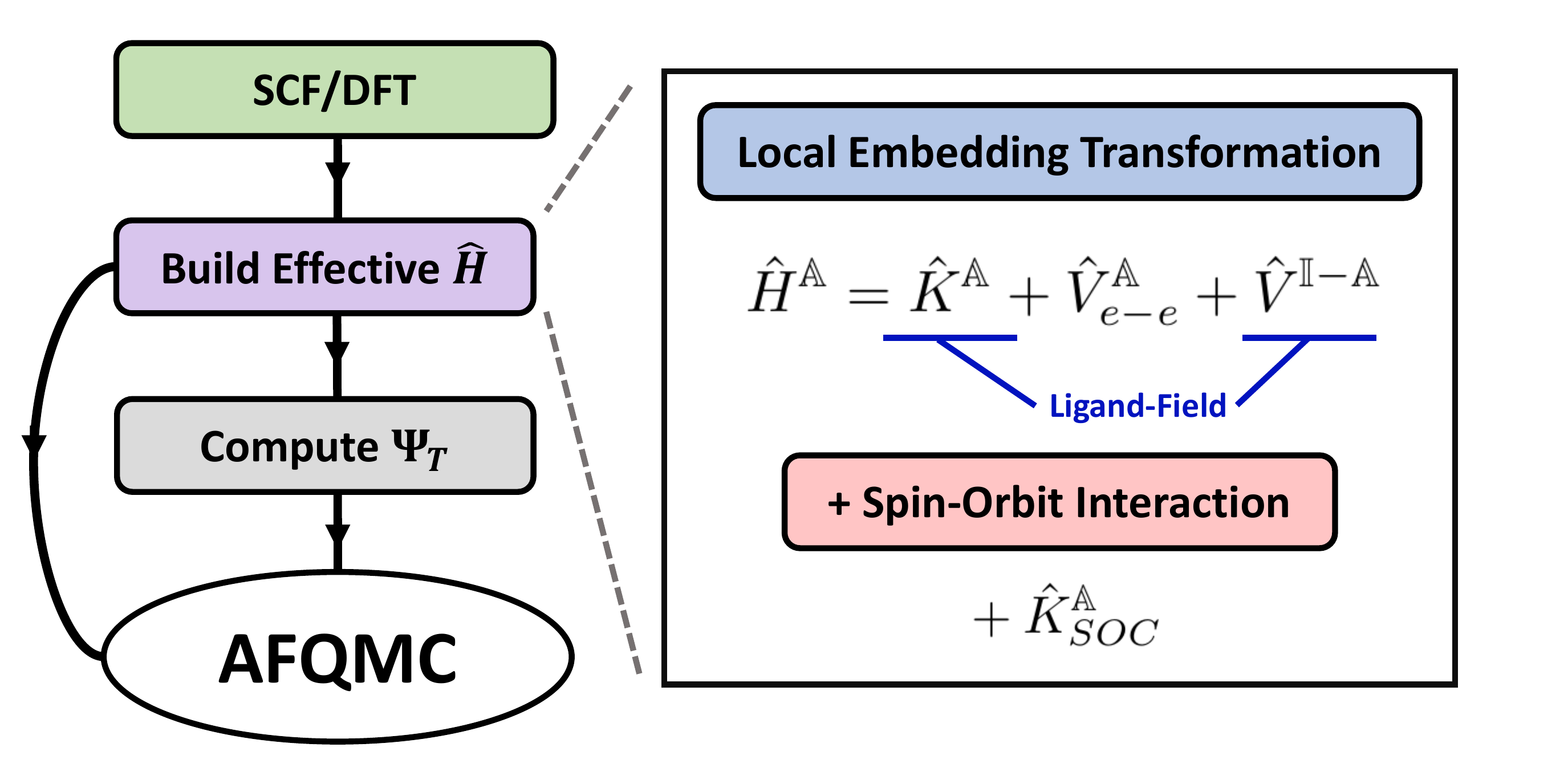}
\caption{
\label{fig:workflow} Schematic of the molecular magnet treatment workflow.
The workflow proceeds from top to bottom.
Boxes represent specific steps in the procedure and 
solid black lines with arrows indicate that the output from one step 
is used as an input to another.
Dotted gray lines indicate a ``breakout'' of the purple box
into two essentially independent substeps.
Each step of the workflow is outlined in the main text.
}
\end{center}
\end{figure}

In this subsection, we describe the treatment of molecular magnets at the many-body level of theory,
including explicit SOC, electron correlation, and ligand field effects.
The basic idea is to produce a 2nd quantized Hamiltonian
from which the low energy spectrum may be computed using AFQMC either with or without SOC.
Figure~\ref{fig:workflow} includes a schematic representation of the 
high-level workflow. 
In the remainder of this section, 
we describe each step in the workflow 
using the schematic as a guide.

Our procedure begins with 
an initial Hartree-Fock or DFT 
calculation, as  indicated in the light green box of Fig.~\ref{fig:workflow}.
This is performed in order to generate an orbital basis, and an electron density, for 
subsequent local embedding 
(DFT$+U$ type calculations can also be used).
This can be accomplished with calculations that neglect SOC 
since typical ligands for molecular magnets consist of light atoms, or atoms which are closed-shell.
For ligands containing heavy elements, such as Br, I, or Bi~\cite{Mn_Bi_molMag_2016, Halide_Anisotropy_2017}, it will 
be useful to include SOC from
the outset.
Here, we used spin restricted open-shell Kohn-Sham (ROKS) DFT 
which provides a convenient starting point for local embedding.
However, spin unrestricted Kohn-Sham (UKS) or generalized Kohn-Sham (GKS) DFT could 
be used instead, if desired.

The next step, 
corresponding to the purple box in Figure~\ref{fig:workflow}, is to construct an effective Hamiltonian for the molecular magnet system.
We require a Hamiltonian which captures material-specific ligand field effects, SOC, and electron correlation.
This is achieved in two steps, as indicated by the dotted breakout section of Fig.~\ref{fig:workflow}.
First, local embedding is employed, as described below, to build an effective Hamiltonian describing the magnetically active
ion(s) in the molecular magnet and which accounts for ligand field effects.
Second,
 a SOC operator, which operates on the same Hilbert space as the effective Hamiltonian,
is constructed and added to the effective Hamiltonian.
We describe both of these steps below.
Of course, 
one can perform calculations with no SOC
by simply neglecting the second step.

We outline the local embedding approach we use in AFQMC~\cite{Purwanto2013,Eskridge2019}.
The crux of local embedding is a separability approximation of the many-body wavefunction,  $\ket{\Psi}$,  into an active and an inactive part,
\eql{eq:separability}
{
\ket{\Psi} \approx \mathcal{A}( \ket{\Psi^\mathbb{I}} \otimes  \ket{\Psi^\mathbb{A}} )
,
}
where $\ket{\Psi^\mathbb{I}}$ is a wavefunction in the inactive space,
$\ket{\Psi^\mathbb{A}}$ is 
the active space many-body wavefunction, and $ \mathcal{A}$ is an antisymmetrizer.
The choice of $\mathbb{A}$ and $\mathbb{I}$ can lead to different forms of embedding, and here they
are chosen based on local criteria.
This approximation allows the energy of  the full Hamiltonian to be mapped onto an effective embedding Hamiltonian: 
\eql{eq:energy_fit}
{
\bra{\Psi} \hat{H} \ket{\Psi} = \bra{\Psi^\mathbb{A}} \hat{H}^\mathbb{A} \ket{\Psi^\mathbb{A}}\,,
}
where $\hat{H}_\mathbb{A}$ is the embedding Hamiltonian which operates only in $\mathbb{A}$ similar to the standard frozen core approximation.
This condition leads to the following explicit form of the embedding Hamiltonian,
\eql{eq:embedding_H}
{
\hat{H}^\mathbb{A}  = \sum_{ij \in \mathbb{A}} K_{ij} \hat{c}^\dagger_i \hat{c}_j + \sum_{ijkl \in \mathbb{A}} V_{ijkl}  \hat{c}^\dagger_i \hat{c}^\dagger_j \hat{c}_l \hat{c}_k + 
 \sum_{ij \in \mathbb{A}} V^{\mathbb{I}-\mathbb{A}}_{ij} \hat{c}^\dagger_i \hat{c}_j + E_\mathbb{I}
 \,,
}
where the first two terms are the one-, and two-body terms in 
 the full Hamiltonian restricted to orbitals within $\mathbb{A}$, 
$E_\mathbb{I}$ is a constant contribution from the energy of the inactive part, 
and $\hat{V}^{\mathbb{I}-\mathbb{A}} = \sum_{ij \in \mathbb{A}} V^{\mathbb{I}-\mathbb{A}}_{ij}  \hat{c}^\dagger_i \hat{c}_j $ is a one-body operator which captures the interaction between active and inactive electrons.
Formally, $\hat{V}^{\mathbb{I}-\mathbb{A}}$
  is an energy consistent, non-local pseudopotential which is computed for the specific system at hand, avoiding transferability errors.
The combination of $\hat{K}$, restricted to $\mathbb{A}$, and 
$\hat{V}^{\mathbb{I}-\mathbb{A}}$
describe 
the symmetry of the full ligand.
Therefore, a small active space focused tightly on a magnetic ion is directly influenced by the ligand field at the many-body level of theory.

The partition between $\mathbb{A}$ and $\mathbb{I}$ is defined in terms of a chosen set of active/inactive local orbitals.
Orbitals are assigned to the active space
 if their centroid is localized within a chosen localization radius of the strongly-correlated center(s), and a separate localization radius is used for occupied orbitals, $R_o$, 
and for virtual orbitals, $R_v$.
The accuracy of the approximation can be dialed up by increasing $R_o$ and $R_v$ and it was previously shown that, for fixed $R_o$, the absolute AFQMC energy converges in $R_v$ at $R_v = R_o + C$ with $C$ being a system-dependent constant typically ranging from 2-6 Bohr~\cite{Eskridge2019}.
However, relative energies, such as the zero-field splitting (ZFS), converge more rapidly, 
allowing for smaller choices of $C$.
In initial test cases, local embedding was observed to reduce the computational cost of some calculations 
by orders of magnitude compared with AFQMC performed on the full Hilbert space.

For the computation of the ZFS gaps, AFQMC calculations are performed with an explicit SOC operator included
in the Hamiltonian.
An explicit spin-orbital basis, $\{\ket{\chi_{\mu}}\}$, is constructed from the set of active local orbitals
corresponding to $\mathbb{A}$.
The second quantized SOC operator can be constructed directly in the spin-orbital basis as 
\eql{eq:active_soc}
{
\Kop^{soc}_{\mu \nu} = \bra{ \chi_\mu }  \hat{\bm{W}}^{soc} \cdot \hat{\bm{S}} \ket{ \chi_\nu }
,
}
which is added to $\hat{H}^\mathbb{A}$ (also transformed to the spin-orbital basis).
Several choices of \textit{ab initio} SOC operators are available in the literature, including all-electron 
and PSP formalisms.
In the case of all-electron relativistic Hamiltonians, the Breit interaction, which is a spin-dependent two-body interaction,  is often modeled 
by an approximate one-body operator. 
This is often done via the spin-orbit mean-field (SOMF) approximation~\cite{SOMF_1996}, in which the full Breit
interaction is replaced with a Fock-like operator constructed from a given electron density,
 although other approximations exist as well~\cite{OneCenterApprox_Breit_1970}.
A detailed discussion of the accuracy of such effective one-body approximations to the Breit interaction is beyond 
the scope of the present work, but 
there are indications from perturbative treatments of molecular magnets~\cite{AccuracyOfSOMF_3dTM}
 that  such approximations provide a reasonable description.
In the case of PSP formalisms, contributions from the Breit interaction are implicitly accounted for,
 again as an effective one-body contribution,  
if the PSP is fit using reference data which accounts for the Breit interaction, 
as is quite common for fully relativistic PSPs.
Equation~\ref{eq:active_soc} is consistent with any effective one-body treatment of SOC, but we adopt the use of relativistic PSPs which
have demonstrated a high degree of accuracy compared with experiment and  
which also have the advantage of allowing SOC to be included selectively for atoms where SOC effects are 
expected to be most import - i.e. for heavy and/or magnetically active ions.

The next step in the workflow, indicated by the light gray box in Figure~\ref{fig:workflow}, is to compute trial wavefunctions for AFQMC.
The embedding Hamiltonian (Eq.~\ref{eq:embedding_H}) is used for this purpose.
(In Sec.~~\ref{sec:results} we perform
calculations both with and without SOC. 
The trial wave function is generated with or without SOC, consistently with the target AFQMC calculation.)
For general excited state calculations, such as the ZFS, targeting the correct quantum numbers and symmetry is as important as the 
the accuracy (as judged by the variational energy, for example).
Many approaches can be used to compute the trial wave function \cite{Shi2021}.
In the present work, we used truncated multi-determinant expansions computed using semistochastic heatbath CI (SHCI)~\cite{Holmes2016, Sharma2017, Sharma2018}, including explicit SOC where needed.
We used  a small active space for SHCI which focuses on the magnetically active electrons.
We note that SHCI can treat much larger active spaces than those used here, 
but that is not needed for the present purpose since the AFQMC results converge quickly with respect to the truncated
 trial wave function.

In molecular magnets, the z-projection of the total angular momentum, $\Jz$, is often an approximately 
good quantum number.
While it is possible to construct rigorous many-body eigenstates of the $\hat{J}_z$ operator for a particular system, 
these may not 
correspond to approximate eigenstates of the Hamiltonian in general and, therefore, may perform poorly as trial wavefunctions for AFQMC. 
Alternatively,
one may utilize a complete set of $\Jz$ eigenstates 
, $\{\ket{\Phi_i^{M_J}}\}$, where $M_J$ is the eigenvalue corresponding to $\hat{J}_z$ and $i$ is an index within the $M_J$ manifold, 
as a basis in which to characterize the approximate $\Jz$ quantum number label via projection.
An arbitrary many-body wavefunction, $\ket{\Psi}$, may be 
expressed as
\eql{eq:MJproj}
{
 \ket{\Psi} = 
  \sum_{M_J = -J}^{J} \sum_{ i \in M_J } C_i^{M_J} \ket{\Phi_i^{M_J}}
\,,
}
where $C_i^{M_J} = \bra{\Phi_i^{M_J}} \Psi \rangle $.
The total weight of $\ket{\Psi}$ which resides within a particular $M_J$-manifold is given by
\eql{eq:MJweight}
{
 W^{M_J} = \sum_{i \in M_J} |C_i^{M_J}|^2
\,.
}
Approximate $M_J$ labels are then assigned based on the 
weights as determined by Eq.~\ref{eq:MJweight}, but only if such an assignment can be made unambiguously.

Our procedure to assign quantum numbers to each trial wavefunction, when possible, is as follows.
While the  dimension of a complete set of $\hat{J}_z$ eigenstates is exponentially large,
the angular momentum is determined by only a handful of $d$-, or $f$-electrons in practice. 
Diagonalizing $\hat{J}_z$ only within the corresponding manifold(s) provides meaningful $M_J$ labels 
while limiting the dimension of the basis of $M_J$ states to a routinely manageable size.
 In this case, the equality in Eq.~\ref{eq:MJproj} no longer holds 
and the accuracy of the approximation can be measured by comparing the total weight of the original wavefunction to 
that of the $M_J$-decomposed wavefunction.
In the present work, 
all SHCI wavefunctions retained an average total weight of 0.9998(1) after being projected 
into the $M_J$ basis.
In the absence of SOC, a similar procedure can be used to assign $M_L$ labels by diagonalizing $\hat{L}_z$ instead of $\hat{J}_z$.
We emphasize that the trial wavefunctions used in AFQMC calculations are 
truncated SHCI wavefunctions, 
which retain only $\mathcal{O}(50)$ determinants;
no attempt was made to force particular quantum numbers in the truncated trial wave function.

In the final step,
AFQMC calculations of the ground state and low-lying excited states are performed using the local embedding Hamiltonian either with (without) 
 SOC, and selecting trial wavefunctions based on their approximate $M_J$ ($M_L$) value.
The approach can be applied to any molecular magnet system including those with several magnetic 
centers (with only minor modifications).
Thus, the AFQMC method provides a general framework for the non-perturbative simulation of molecular magnets.
We demonstrate this framework in Sec.~\ref{sec:results} below.

\section{Application to a linear Co$^{2+}$ complex}
\label{sec:results}

The \CoIIcomplex molecule 
 was recently synthesized and experimentally characterized, 
displaying magnetic hysteresis at temperatures of up to 5K~\cite{CoII_linear_complex}.
It is, to our knowledge, the current record holder for ZFS gap among single ion molecular magnets
based on 3d transition metals.
The large ZFS gap 
 is due to unquenched orbital angular momentum in the ground state, 
which is unusual for $3d$-element complexes.
The Co$^{2+}$ ion at the core of the \CoIIcomplex molecule has similar electronic structure to 
a Co$^{2+}$ ion adsorbed on the surface of MgO~\cite{CoII_on_MgO}, which also displays unquenched orbital angular momentum in the ground state.

The weak $S_6$ ligand field 
 and the locally-linear coordination environment of the $\textrm{Co}^{2+}$ ion 
 at the center of the \CoIIcomplex molecule 
 lead to a $C_{\infty v}$ pseduosymmetry which provides approximate symmetry 
 labels  each corresponding to a well defined eigenvalue of $\Lz$.
Neglecting SOC, a Co$^{2+}$ ion in vacuum has a $^4\textrm{F}$ ground state.
Under a $C_{\infty v}$ ligand field, the \LSterm{4}{F} state is split 
 into $^{4}\Sigma$, $^{4}\Pi$, $^{4}\Delta$, and $^{4}\Phi$ where each level is two-fold degenerate in orbital degrees of freedom except for $^{4}\Sigma$, which is non-degenerate.
Even a modestly strong 
 $C_{\infty v}$ field would typically lead to a $^4\Sigma$ ground state; 
 however, the weak ligand field in \CoIIcomplex 
leads to a $^4\Phi$ ground state instead.
If SOC is included, 
$\hat{L}_z$ no longer provides a good quantum number and
 the $^4\Phi$ state is split into 
 eigenstates  of $\Jz$, which range from $M_J=9/2$ to $M_J=3/2$. 
The ground state of \CoIIcomplex has $M_J=9/2$ as determined by DC magnetic susceptibility data from the literature, 
and the first excitation ZFS gap of 450 $cm^{-1}$ is attributed to an excitation to the $M_J=7/2$ level.
Since Co$^{2+}$ is a Kramer's ion, \CoIIcomplex has exact two-fold degeneracy regardless of the ligand field symmetry.
In our discussions below, $C_{\infty v}$ and $M_J$ labels are only approximate and are determined as
described in Sec.~\ref{sec:molmag_treatment}.

We apply the general computational framework 
 described in Sec.~\ref{sec:molmag_treatment} to 
 compute the low-energy spectrum of \CoIIcomplex, both with and without SOC.
ZFS gaps are taken directly from the low energy spectrum computed with SOC.
The Co ion is treated with 
the CRENBL PSP 
 (which  is based on 
 fully relativistic reference data), 
using 
 the corresponding uncontracted Gaussian primitive basis~\cite{CRENBL_ECP}.
All other atoms are treated 
 with the non-relativistic all-electron Hamiltonian 
 using the standard cc-pVDZ basis for C, O, and Si and the STO-6G basis for H.
We verified 
that the cc-pVDZ basis for ligand atoms is adequate for the calculations performed here.

\begin{figure}
\begin{center}
\includegraphics[width=0.95\textwidth]{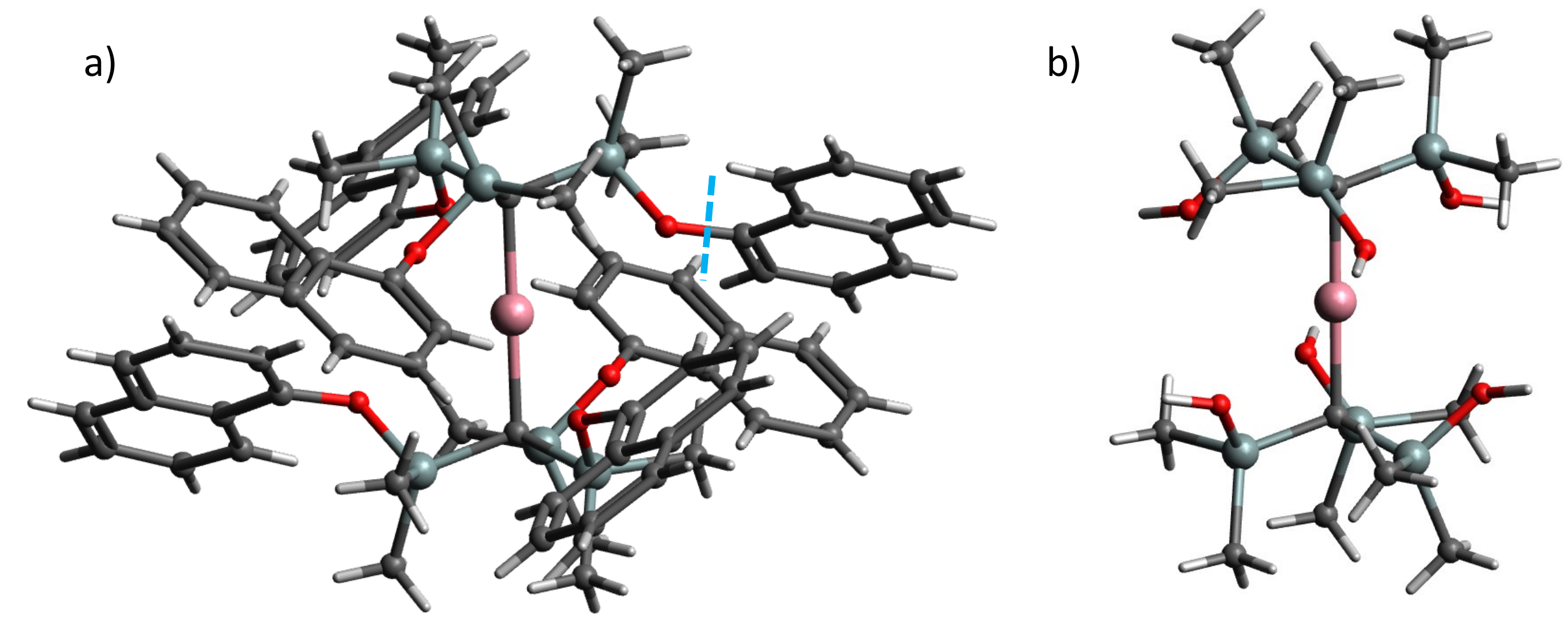}
\caption{
\label{fig:molecule} 
Molecular geometry of the \CoIIcomplex molecule.
Atomic species are identified by color.
Co atoms are pink, Si are pale green, O are red, C are dark gray, and H are light gray/white.
Panel a) shows the full molecule as determined by x-ray diffraction experiments from the literature~\cite{CoII_linear_complex}.
The molecule has $\textrm{S}_6$ symmetry, and the central C-Co-C axis is 
approximately linear.
Panel b) shows the simplified model used in the present work.
The six naphthol ($\mathrm{C}_{10} \mathrm{H}_7$) units in panel a) are replaced with an H-termination,  
holding the Si-O-H angle unchanged.
One of the cut O-C bonds is indicated by a light blue dotted line in panel a).
}
\end{center}
\end{figure}

The experimentally observed geometry 
 of the Co(C(SiMe$_2$ONaph)$_3$)$_2$ molecule~\cite{CoII_linear_complex} is shown in panel a) of Fig.~\ref{fig:molecule}.
The ligand has $S_6$ symmetry; however, all calculations were performed without imposing point-group symmetry.
To assist in converging the initial DFT calculations, we 
replaced each of the six naphthol (``Naph'' = $C_{10}H_7$) units with a hydrogen termination using an O-H bondlength of 1.04 \r{A} and maintaining the original Si-O-C bond angle;
the simplified geometry, \CoIImodel, is illustrated in panel b) of Fig.~\ref{fig:molecule}.
No further geometry optimization was performed on \CoIImodel.
Results in the literature showed that the ZFS computed for \CoIIcomplex and the ZFS computed for a model in which napthol units were replaced
by a methyl group were essentially identical level-by-level, with a maximum deviation of 14 $cm^{-1}$ but with most levels agreeing to within 3 $cm^{-1}$~\cite{CoII_linear_complex}.
This suggests that 
the ZFS is not sensitive to the details of the ligand for this particular complex, 
which is unsurprising given 
the very weak ligand field strength.

Initial DFT calculations were performed using the PBE0 functional and including only 
 the scalar relativistic part of the Co PSP with no SOC.
A local embedding Hamiltonian was constructed as described in Sec.~\ref{sec:molmag_treatment} using the PBE0 solution, 
Foster-Boys localized~\cite{Boys1960} 
restricted open-shell Kohn-Sham (ROKS) 
orbitals as a basis, and
localization radii $(R_o, R_v) = (2.8, 5.4) $ atomic units centered at the Co ion.
This choice of localization radii yields an active space which includes
all Co occupied and virtual orbitals, and 
 some ligand orbitals, for a total of 99 spatial orbitals, or 198 spin-orbitals.
The $3s$, $3p$ and $3d$ electrons belonging to the Co$^{2+}$ ion are all included
 in the active space, as well as a total of 4 additional electrons from neighboring C atoms, for a total of 19 active electrons.
We checked that the choice of ($R_o$, $R_v$) is sufficient for the purpose of computing the ZFS levels 
in \CoIImodel.

We computed trial wavefunctions for AFQMC using SHCI 
as implemented in the code ``Dice''~\cite{Holmes2016, Sharma2017, Sharma2018}.
We performed ROHF on the embedding Hamiltonian with no SOC to provide a reference determinant for SHCI, and to define the Co $3d$-orbitals.
SHCI calculations, both with and without SOC, were performed for 
a small active space consisting of 7 orbitals and 11 electrons, which include both the $3d$ orbitals/electrons
and some ligand orbitals/electrons; in the SHCI,
a variational cutoff of 1.0E-5 was used.
We then truncate the SHCI wavefunction by discarding 
determinants with small weights 
using a truncation threshold of 0.001, to obtain the trial wave functions for AFQMC.
We assigned labels to each trial wavefunction corresponding to $\hat{J}_z$ and $\hat{L}_z$
for calculations performed with SOC and without SOC, respectively, 
 as described in Sec.~\ref{sec:molmag_treatment}.
We confirmed that AFQMC maintains the same $M_J$ ($M_L$) labels as the trial wavefunctions. 

\begin{figure}
\begin{center}
\includegraphics[width=1.0\textwidth]{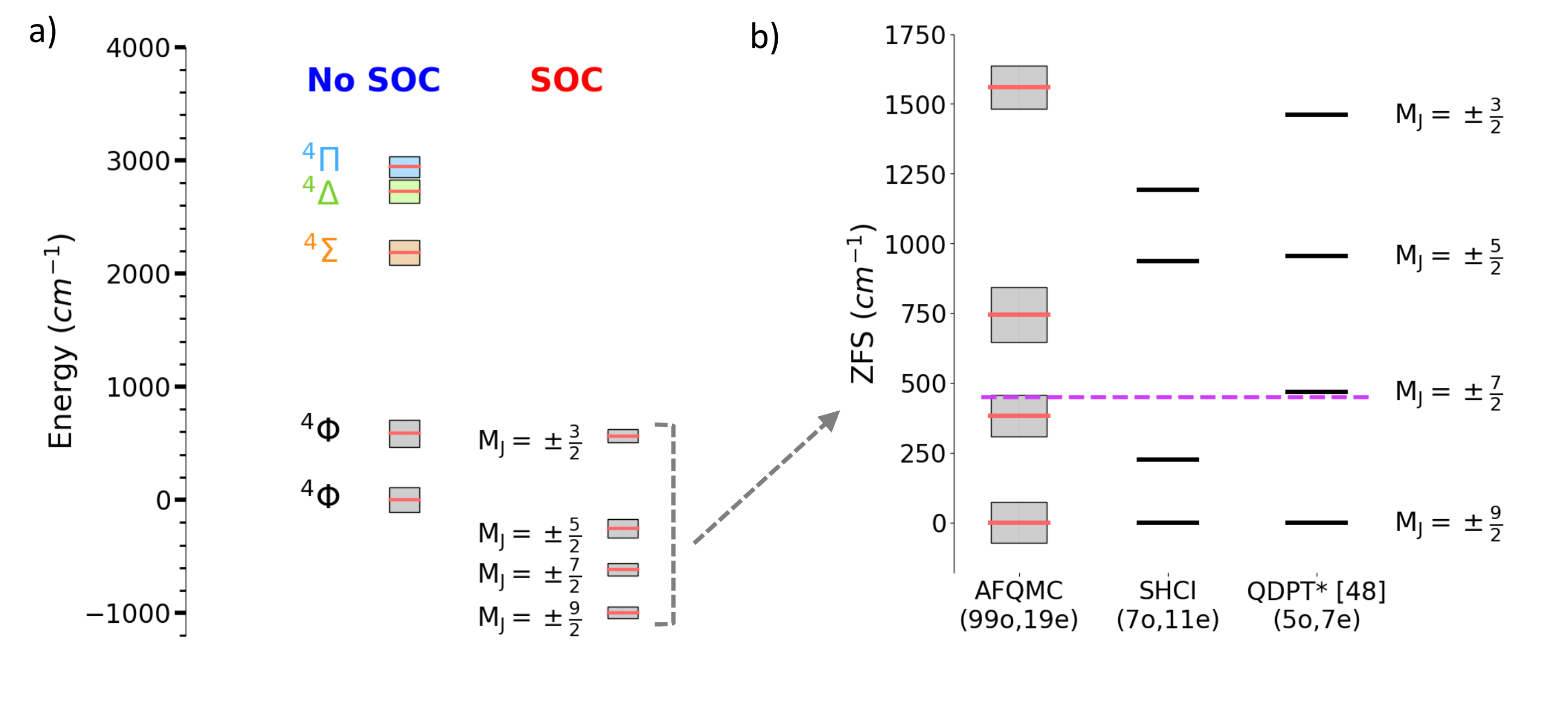}
\caption{
\label{fig:ZFS_afqmc}
AFQMC results for the \CoIImodel molecule.
Panel a) shows 
energy levels with (right column) and without (left column) SOC.
Red horizontal lines are AFQMC energies and shaded
 boxes are the stochastic uncertainty.
Colored boxes in the entire figure correspond to the approximate $C_{\infty v}$ labels of ``No SOC''.
Approximate symmetry labels (see main text) 
are displayed next to the corresponding AFQMC energy level.
The AFQMC energies 
are plotted on the same absolute scale using the ``No SOC'' $^4 \Phi$ ground state energy  
($ -2881.97934\,E_{Ha}$) as a reference.
Panel b) contains the ZFS computed with AFQMC, 
corresponding to the ``SOC'' column in panel a).
For reference, 
the ZFS
computed in the SHCI(7o,11e) calculations, from which we obtain a truncated trial wavefunction  for AFQMC, is also shown, 
together with the result from SA-CASSCF(5o,7e) + QDPT \cite{CoII_linear_complex}  (labelled ``QDPT* (5o,7e)'').
The experimentally observed ZFS is plotted as a purple horizontal dotted line.
}
\end{center}
\end{figure}

The low energy many-body spectrum of  \CoIImodel computed from AFQMC
 is shown in panel a) of Fig.~\ref{fig:ZFS_afqmc}.
Results are organized into two columns.
The left column, labelled ``No SOC'', is for a Hamiltonian 
with only the scalar relativistic PSP of Co, 
without SOC. 
The AFQMC ground state from the ``No SOC'' column is used as an absolute energy reference for
all of panel a).
Symmetry labels in the ``No SOC'' column correspond to approximate C$_{\infty v}$ labels.
The right column, labelled ``SOC'', is for a Hamiltonian 
which contains both the scalar relativistic and  the SOC parts 
of the Co PSP.
In all panels of Fig.~\ref{fig:ZFS_afqmc}, 
red horizontal lines are AFQMC
 energy levels, 
  and boxes indicate the 
  stochastic uncertainties.
 Besides AFQMC results, we also show the results of the modest SHCI calculation from which the trial wavefunction was generated, as well as 
 a  SA-CASSCF(5o,7e) + QDPT result from the literature~\cite{CoII_linear_complex}, for reference. 
 The experimentally measured ZFS~\cite{CoII_linear_complex} of 450 $cm^{-1}$, determined to 
 be from the gap between an $M_J=9/2$ ground state and $M_J=7/2$ first excited state,
 is also included in panel b). 

The ``No SOC'' spectrum in Fig.~\ref{fig:ZFS_afqmc} corresponds to all levels that nominally originate from the $^4\textrm{F}$ ground state of a Co$^{2+}$ ion in vacuum split by a $C_{\infty v}$ ligand field.
The ground state is a $^4\Phi$ state, with unquenched orbital angular momentum, which would be doubly degenerate under perfect $C_{\infty v}$ symmetry; however, the system has only $S_6$ symmetry.
A small gap of 590(178) $cm^{-1}$ exists between the ground state and its approximate $^4\Phi$ pair.
Nonrelativistic state-averaged CASSCF(5o,7e) calculations order the states as $^4\Phi$, $^4\Sigma$, $^4\Delta$, then $^4\Pi$ \cite{CoII_linear_complex}, similar to our AFQMC results.
The $^4\Delta$ and $^4\Pi$ states computed with AFQMC are  
separated by only 217(138)\,cm$^{-1}$.
The total spread of the splitting of a hypothetical $^4$F state under the ligand field in \CoIImodel 
is 2717(281)\,cm$^{-1}$, similar to the estimate of  2768\,cm$^{-1}$  obtained from non-relativistic SA-CASSCF(5o,7e)\,+\,NEVPT2~\cite{CoII_linear_complex}. 
The relatively small spread 
reflects the weak ligand field strength.

In Fig.~\ref{fig:ZFS_afqmc} the low energy many-body spectrum of \CoIImodel, with SOC included,
is shown in the right column of panel a).
The AFQMC calculations were run longer for 
``SOC'' than for 
``No SOC'' (beyond the expected increase in computational demand due to the doubling of the Slater determinants), 
in order to gather more statistics to resolve the small energy scales.
All states which originate from both of the 
 $^4\Phi$ states in the ``No SOC'' column are included.
All levels computed with SOC are two-fold degenerate due to time reversal symmetry,
and AFQMC was run for only 
one of the two states. 
AFQMC predicts an $M_J=9/2$ ground state, consistent with experiment.
The effect from SOC is seen to contribute on roughly the same energy scale as the splittings induced by the ligand field.

Panel b) in Fig.~\ref{fig:ZFS_afqmc}  shows a zoomed in view of the computed  ZFS, with respect to the energy of the ground state, $M_J=9/2$.
The ZFS computed using SHCI(7o,11e) correspond to the calculations used to obtain trial wavefunctions for AFQMC and 
are derived from the variational energy of the SHCI wavefunctions.
For comparison, we inlcuded
the ZFS 
computed using SA-CASSCF(5o,7e) + QDPT, as well as  experimental values, all taken from Ref.~~\cite{CoII_linear_complex}.
The experimental result, displayed as a purple horizontal dotted line, is from variable-field far-infrared spectroscopy, covering an energy range of 30-600 $cm^{-1}$,
and represents a direct measure of the ZFS.
We note that the energy range explored by experiment does not reach the next gap predicted by either AFQMC or by SA-CASSCF(5o,7e)+QDPT.
The gap between the $M_J = 9/2$ ground state and the $M_J = 7/2$ state
 computed by AFQMC is 382(74)\,cm$^{-1}$, where the error bar is the joint  
 statistical uncertainty of the two separate calculations. 
This agrees with the experimentally observed value of 450\,cm$^{-1}$. 
SA-CASSCF(5o,7e)\,+\,QDPT 
 yields an excitation gap of 468 $cm^{-1}$, which agrees remarkably well with experiment.
The spectrum computed from AFQMC provides a useful benchmark.
It is interesting to note that, in addition to the first excitation,  SA-CASSCF(5o,7e)\,+\,QDPT also yields a spectrum 
in good agreement with AFQMC,  over-estimating the second excitation slightly.

\section{Conclusion and Outlook}
\label{sec:Discussion}
\label{sec:Conclusion}

We have developed a general non-perturbative approach to treat molecular magnets using AFQMC, in which material specificity, static and dynamic electron correlation, and SOC are treated accurately and on an equal footing.
 As a first application, we test the method on 
 a slightly simplified model 
  of the \CoIIcomplex molecule.
  {\it Ab initio\/} AFQMC incorporating SOC and treating the interaction in a large active space  
   yields a ZFS gap of 382(74)\,cm$^{-1}$, between the $M_J=9/2$ ground state and the $M_J=7/2$ excited state, consistent with 
the experimental value of 450\,cm$^{-1}$.
The computed low-lying excitations can serve as a useful benchmark for future work in this system and for method 
developments. 
    
We expect this approach to be applicable to many other related systems for high accuracy, predictive calculations. 
A number of methodological improvements can be made to further improve the 
computational efficiency, in order to treat even larger systems or target higher statistical accuracies.
The \CoIIcomplex molecule has one of the largest known ZFS gaps of 3d-transition metal based molecular magnets.
Most others have ZFS gaps which are smaller by at least a factor of 2-3, often more.
Here, we computed the ZFS by performing independent AFQMC calculations for each $M_J$ level;
Sufficient resolution was achieved with this approach.
For computing the ZFS for general molecular magnets, a higher resolution can be reached via 
correlated sampling, 
which directly computes relative energies,
such as the ZFS gaps, 
 with significantly reduced Monte Carlo noise~\cite{CorrelatedSampling2017}.

We computed the low-energy many-body spectrum of the \CoIImodel molecule.
AFQMC can also be used to directly compute general observables, such as forces, electron density, and phonon/vibrational modes, via the back-propagation algorithm~\cite{Purwanto2004,BackProp2017}.
The use of back-propagation 
 would follow essentially the same procedure as shown in Fig.~\ref{fig:workflow}
  where the final step would be AFQMC with back-propagation instead of the energy-only calculations
 performed here.
 This is, of course, highly desirable since 
structural details (e.g.~ligand symmetry) and vibrational properties are key elements of molecular magnet design.
Much of  the molecular magnet literature relies upon the experimentally measured geometry of molecular magnets in order to perform ZFS calculations.
AFQMC offers a direct a route to the \textit{ab initio} prediction of equilibrium geometries as has been demonstrated in solids~\cite{arXiv:2302.07460} 
which would greatly assist in the design of new molecular magnets.

Molecular magnets are very large both in terms of their spatial extent, and the dimension of the corresponding many-body Hilbert space.
In the procedure described in Sec.~\ref{sec:molmag_treatment}, local embedding AFQMC~\cite{Eskridge2019} is employed 
as a way to reduce the size of the active space 
 while retaining system-specific details.
In the case of \CoIImodel, local embedding led to a reduction in computational cost by a factor of approximately 5000 relative to a hypothetical AFQMC calculation performed using the full Hilbert space.
Local embedding benefits greatly from a cancellation of errors in 
 relative energy calculations, such as for the ZFS, and is systematically improvable 
 towards full AFQMC by increasing the localization radii ($R_o$, $R_v$).
Direct AFQMC computations of the full 
molecular magnets without embedding
is also rapidly becoming feasible,
by taking advantage of GPU-acceleration~\cite{GPU_shee_2018, GPU_solids_2020} and efficient multi-determinant trial wavefunction algorithms~\cite{Shi2021,FastSHCItrial}.
The combination of effective embedding approaches with these advances in computational efficiency 
will spur a large number of applications in molecular magnets and beyond.



\acknowledgements

We thank Kyungwha Park, and James Shee for helpful discussions. 
We also acknowledge support from the U.S. Department
of Energy (DOE) under grant DE-SC0001303.
The Flatiron Institute is a division of the Simons Foundation.
The authors acknowledge William \& Mary Research Computing 
for providing computational resources and/or technical support that
 have contributed to the results reported within this paper.
URL: https://www.wm.edu/it/rc




\bibliographystyle{phaip.bst}
\bibliography{manuscript.bib}

\begin{thebibliography}{10}

\bibitem{Mn12MolMag1993}
R.~Sessoli, D.~Gatteschi, A.~Caneschi, and M.~A. Novak,
\newblock Nature {\bf 365}, 141 (1993).

\bibitem{Vincent2012}
R.~Vincent, S.~Klyatskaya, M.~Ruben, W.~Wernsdorfer, and F.~Balestro,
\newblock Nature {\bf 488}, 357 (2012).

\bibitem{PhysRevB.87.195412}
M.~Urdampilleta, S.~Klyatskaya, M.~Ruben, and W.~Wernsdorfer,
\newblock Phys. Rev. B {\bf 87}, 195412 (2013).

\bibitem{TbPc2_electrically_driven}
S.~Thiele et~al.,
\newblock Science {\bf 344}, 1135 (2014).

\bibitem{TbPc2_Grovers_2017}
C.~Godfrin et~al.,
\newblock Phys. Rev. Lett. {\bf 119}, 187702 (2017).

\bibitem{PerspectiveMolMag_Qinfo}
A.~Gaita-Ari{\~{n}}o, F.~Luis, and E.~Hill, S.and~Coronado,
\newblock Nature Chemistry {\bf 11}, 301 (2019).

\bibitem{Cr4_opticallyAddressable}
S.~L. Bayliss et~al.,
\newblock Science {\bf 370}, 1309 (2020).

\bibitem{D2DT01440H}
J.~L.~S. Milani et~al.,
\newblock Dalton Trans. {\bf 51}, 12258 (2022).

\bibitem{FeII_molMag_2010}
D.~E. Freedman et~al.,
\newblock Journal of the American Chemical Society {\bf 132}, 1224 (2010),
\newblock PMID: 20055389.

\bibitem{family_of_FeII_molMags_2010}
W.~H. Harman et~al.,
\newblock Journal of the American Chemical Society {\bf 132}, 18115 (2010),
\newblock PMID: 21141856.

\bibitem{FeII_molMag_2011}
D.~Weismann et~al.,
\newblock Chemistry – A European Journal {\bf 17}, 4700 (2011).

\bibitem{CoII_molMag_2011}
J.~M. Zadrozny and J.~R. Long,
\newblock Journal of the American Chemical Society {\bf 133}, 20732 (2011),
\newblock PMID: 22142241.

\bibitem{Linear_FeI_molMag_2013}
J.~M. Zadrozny et~al.,
\newblock Nature Chemistry {\bf 5}, 577 (2013).

\bibitem{FeII_molMag_2013}
J.~M. Zadrozny et~al.,
\newblock Chem. Sci. {\bf 4}, 125 (2013).

\bibitem{Co_molMag_2014}
M.~S. Fataftah, J.~M. Zadrozny, D.~M. Rogers, and D.~E. Freedman,
\newblock Inorganic Chemistry {\bf 53}, 10716 (2014),
\newblock PMID: 25198379.

\bibitem{NiII_molMag_2015}
K.~E.~R. Marriott et~al.,
\newblock Chem. Sci. {\bf 6}, 6823 (2015).

\bibitem{TbPc2_Original_2003}
N.~Ishikawa, M.~Sugita, T.~Ishikawa, S.-y. Koshihara, and Y.~Kaizu,
\newblock Journal of the American Chemical Society {\bf 125}, 8694 (2003),
\newblock PMID: 12862442.

\bibitem{doi:10.1021/jacs.6b02638}
J.~Liu et~al.,
\newblock Journal of the American Chemical Society {\bf 138}, 5441 (2016),
\newblock PMID: 27054904.

\bibitem{doi:10.1021/jacs.5b13584}
Y.-C. Chen et~al.,
\newblock Journal of the American Chemical Society {\bf 138}, 2829 (2016),
\newblock PMID: 26883386.

\bibitem{DyMetallocene2018}
F.-S. Guo et~al.,
\newblock Science {\bf 362}, 1400 (2018).

\bibitem{ActinideMolMagRev2015}
K.~R. Meihaus and J.~R. Long,
\newblock Dalton Trans. {\bf 44}, 2517 (2015).

\bibitem{doi:10.1021/ja906012u}
J.~D. Rinehart and J.~R. Long,
\newblock Journal of the American Chemical Society {\bf 131}, 12558 (2009),
\newblock PMID: 19689136.

\bibitem{Neptunocene}
N.~Magnani et~al.,
\newblock Angewandte Chemie International Edition {\bf 50}, 1696 (2011).

\bibitem{Pu_MolMag_2018}
C.~A. Gaggioli and L.~Gagliardi,
\newblock Inorganic Chemistry {\bf 57}, 8098 (2018),
\newblock PMID: 29968473.

\bibitem{WhatIsNotRequired4MolMags}
F.~Neese and D.~A. Pantazis,
\newblock Faraday Discuss. {\bf 148}, 229 (2011).

\bibitem{C4CS00439F}
G.~A. Craig and M.~Murrie,
\newblock Chem. Soc. Rev. {\bf 44}, 2135 (2015).

\bibitem{Toward_HighT_SMMs}
L.~Ungur and L.~F. Chibotaru,
\newblock Inorganic Chemistry {\bf 55}, 10043 (2016),
\newblock PMID: 27508399.

\bibitem{Review_molMag_2019}
F.-S. Guo, A.~K. Bar, and R.~A. Layfield,
\newblock Chemical Reviews {\bf 119}, 8479 (2019),
\newblock PMID: 31059235.

\bibitem{Hait2019}
D.~Hait, N.~M. Tubman, D.~S. Levine, K.~B. Whaley, and M.~Head-Gordon,
\newblock Journal of Chemical Theory and Computation {\bf 15}, 5370 (2019),
\newblock PMID: 31465217.

\bibitem{Shee2021}
J.~Shee, M.~Loipersberger, D.~Hait, J.~Lee, and M.~Head-Gordon,
\newblock The Journal of Chemical Physics {\bf 154}, 194109 (2021).

\bibitem{SACASSCF_1980}
B.~O. Roos, P.~R. Taylor, and P.~E. Sigbahn,
\newblock Chemical Physics {\bf 48}, 157 (1980).

\bibitem{SACASSCF_1981}
P.~E.~M. Siegbahn, J.~Almlöf, A.~Heiberg, and B.~O. Roos,
\newblock The Journal of Chemical Physics {\bf 74}, 2384 (1981).

\bibitem{NEVPT2_2001}
C.~Angeli, R.~Cimiraglia, S.~Evangelisti, T.~Leininger, and J.-P. Malrieu,
\newblock The Journal of Chemical Physics {\bf 114}, 10252 (2001).

\bibitem{NEVPT2_2002}
C.~Angeli, R.~Cimiraglia, and J.-P. Malrieu,
\newblock The Journal of Chemical Physics {\bf 117}, 9138 (2002).

\bibitem{CASPT2_1990}
K.~Andersson, P.~A. Malmqvist, B.~O. Roos, A.~J. Sadlej, and K.~Wolinski,
\newblock The Journal of Physical Chemistry {\bf 94}, 5483 (1990).

\bibitem{CASPT2_1991}
K.~Andersson, P.~Malmqvist, and B.~O. Roos,
\newblock The Journal of Chemical Physics {\bf 96}, 1218 (1992).

\bibitem{msCASPT2_1998}
J.~Finley, P.~Åke Malmqvist, B.~O. Roos, and L.~Serrano-Andrés,
\newblock Chemical Physics Letters {\bf 288}, 299 (1998).

\bibitem{QDPT_2006}
D.~Ganyushin and F.~Neese,
\newblock The Journal of Chemical Physics {\bf 125}, 024103 (2006).

\bibitem{RASSI_2002}
P.~Åke Malmqvist, B.~O. Roos, and B.~Schimmelpfennig,
\newblock Chemical Physics Letters {\bf 357}, 230 (2002).

\bibitem{Zhang2003}
S.~Zhang and H.~Krakauer,
\newblock Phys. Rev. Lett. {\bf 90}, 136401 (2003).

\bibitem{AlSaidi2006b}
W.~A. Al-Saidi, S.~Zhang, and H.~Krakauer,
\newblock J. Chem. Phys. {\bf 124}, 224101 (2006).

\bibitem{Eskridge2022}
B.~Eskridge, H.~Krakauer, H.~Shi, and S.~Zhang,
\newblock The Journal of Chemical Physics {\bf 156}, 014107 (2022).

\bibitem{DirectCompMB2020}
K.~T. Williams et~al.,
\newblock Phys. Rev. X {\bf 10}, 011041 (2020).

\bibitem{TM_bench_2019}
J.~Shee et~al.,
\newblock Journal of Chemical Theory and Computation {\bf 15}, 2346 (2019),
\newblock PMID: 30883110.

\bibitem{TM_bench_2020}
B.~Rudshteyn et~al.,
\newblock Journal of Chemical Theory and Computation {\bf 16}, 3041 (2020),
\newblock PMID: 32293882.

\bibitem{TM_metallocene_2022}
B.~Rudshteyn et~al.,
\newblock Journal of Chemical Theory and Computation {\bf 18}, 2845 (2022),
\newblock PMID: 35377642.

\bibitem{Eskridge2019}
B.~Eskridge, H.~Krakauer, and S.~Zhang,
\newblock Journal of Chemical Theory and Computation {\bf 15}, 3949 (2019),
\newblock PMID: 31244125.

\bibitem{CoII_linear_complex}
P.~C. Bunting et~al.,
\newblock Science {\bf 362}, eaat7319 (2018).

\bibitem{Motta2018}
M.~Motta and S.~Zhang,
\newblock WIREs Computational Molecular Science {\bf 8}, e1364 (2018).

\bibitem{Shi2021}
H.~Shi and S.~Zhang,
\newblock The Journal of Chemical Physics {\bf 154}, 024107 (2021).

\bibitem{THOULESS1960225}
D.~Thouless,
\newblock Nuclear Physics {\bf 21}, 225 (1960).

\bibitem{Purwanto2011}
W.~Purwanto, H.~Krakauer, Y.~Virgus, and S.~Zhang,
\newblock J. Chem. Phys. {\bf 135}, 164105 (2011).

\bibitem{Trotter1959}
H.~F. Trotter,
\newblock Proc.\ Am.\ Math.\ Soc. {\bf 10}, 545 (1959).

\bibitem{Suzuki1976}
M.~Suzuki,
\newblock Commun.\ Math.\ Phys. {\bf 51}, 183 (1976).

\bibitem{Stratonovich1957}
R.~D. Stratonovich,
\newblock Dokl.\ Akad.\ Nauk.\ SSSR {\bf 115}, 1907 (1957).

\bibitem{Hubbard1959}
J.~Hubbard,
\newblock Phys. Rev. Lett. {\bf 3}, 77 (1959).

\bibitem{Zhang2019}
S.~Zhang,
\newblock Auxiliary-field quantum monte carlo at zero- and finite-temperatures,
\newblock edited by E. Pavarini, E. Koch, and S. Zhang (Verlag des
  Forschungszentrum J\"{u}lich, 2019), Vol. \textbf{9}.

\bibitem{Purwanto2009_C2}
W.~Purwanto, S.~Zhang, and H.~Krakauer,
\newblock J. Chem. Phys. {\bf 130}, 094107 (2009).

\bibitem{Ma_2013}
F.~Ma, S.~Zhang, and H.~Krakauer,
\newblock New Journal of Physics {\bf 15}, 093017 (2013).

\bibitem{Mn_Bi_molMag_2016}
T.~J. Pearson, M.~S. Fataftah, and D.~E. Freedman,
\newblock Chem. Commun. {\bf 52}, 11394 (2016).

\bibitem{Halide_Anisotropy_2017}
S.~C. Coste, B.~Vlaisavljevich, and D.~E. Freedman,
\newblock Inorganic Chemistry {\bf 56}, 8195 (2017),
\newblock PMID: 28661134.

\bibitem{Purwanto2013}
W.~Purwanto, S.~Zhang, and H.~Krakauer,
\newblock Journal of Chemical Theory and Computation {\bf 9}, 4825 (2013),
\newblock PMID: 26583401.

\bibitem{SOMF_1996}
B.~A. Heß, C.~M. Marian, U.~Wahlgren, and O.~Gropen,
\newblock Chemical Physics Letters {\bf 251}, 365 (1996).

\bibitem{OneCenterApprox_Breit_1970}
T.~E.~H. Walker and W.~G. Richards,
\newblock The Journal of Chemical Physics {\bf 52}, 1311 (1970).

\bibitem{AccuracyOfSOMF_3dTM}
J.~Netz, A.~O. Mitrushchenkov, and A.~Köhn,
\newblock Journal of Chemical Theory and Computation {\bf 17}, 5530 (2021),
\newblock PMID: 34388346.

\bibitem{Holmes2016}
A.~A. Holmes, N.~M. Tubman, and C.~J. Umrigar,
\newblock Journal of Chemical Theory and Computation {\bf 12}, 3674 (2016),
\newblock PMID: 27428771.

\bibitem{Sharma2017}
S.~Sharma, A.~A. Holmes, G.~Jeanmairet, A.~Alavi, and C.~J. Umrigar,
\newblock Journal of Chemical Theory and Computation {\bf 13}, 1595 (2017),
\newblock PMID: 28263594.

\bibitem{Sharma2018}
B.~Mussard and S.~Sharma,
\newblock Journal of Chemical Theory and Computation {\bf 14}, 154 (2018),
\newblock PMID: 29202220.

\bibitem{CoII_on_MgO}
I.~G. Rau et~al.,
\newblock Science {\bf 344}, 988 (2014).

\bibitem{CRENBL_ECP}
M.~M. Hurley, L.~F. Pacios, P.~A. Christiansen, R.~B. Ross, and W.~C. Ermler,
\newblock The Journal of Chemical Physics {\bf 84}, 6840 (1986).

\bibitem{Boys1960}
S.~F. Boys,
\newblock Rev. Mod. Phys. {\bf 32}, 296 (1960).

\bibitem{CorrelatedSampling2017}
J.~Shee, S.~Zhang, D.~R. Reichman, and R.~A. Friesner,
\newblock Journal of Chemical Theory and Computation {\bf 13}, 2667 (2017),
\newblock PMID: 28481546.

\bibitem{Purwanto2004}
W.~Purwanto and S.~Zhang,
\newblock Phys. Rev. E {\bf 70}, 056702 (2004).

\bibitem{BackProp2017}
M.~Motta and S.~Zhang,
\newblock Journal of Chemical Theory and Computation {\bf 13}, 5367 (2017),
\newblock PMID: 29053270.

\bibitem{arXiv:2302.07460}
S.~Chen and S.~Zhang,
\newblock Computation of forces and stresses in solids: towards accurate
  structural optimizations with auxiliary-field quantum monte carlo,
\newblock \url{https://arxiv.org/abs/2302.07460}, 2023.

\bibitem{GPU_shee_2018}
J.~Shee, E.~J. Arthur, S.~Zhang, D.~R. Reichman, and R.~A. Friesner,
\newblock Journal of Chemical Theory and Computation {\bf 14}, 4109 (2018),
\newblock PMID: 29897748.

\bibitem{GPU_solids_2020}
F.~D. Malone, S.~Zhang, and M.~A. Morales,
\newblock Journal of Chemical Theory and Computation {\bf 16}, 4286 (2020),
\newblock PMID: 32437147.

\bibitem{FastSHCItrial}
A.~Mahajan, J.~Lee, and S.~Sharma,
\newblock The Journal of Chemical Physics {\bf 156}, 174111 (2022).

\end{thebibliography}

\end{document}